\documentclass[manuscript]{emulateapj}
\usepackage{natbib}
\usepackage{graphicx}
\usepackage{txfonts}

\makeatletter

\newcommand{\ser}{S\'ersic }

\slugcomment{Accepted for publication in ApJ}
\shorttitle{The Fundamental Plane of massive quiescent galaxies out to $z\sim2$}
\shortauthors{van de Sande et al.}

\begin{document}

\title{The Fundamental Plane of massive quiescent galaxies out to $z\sim2$}

\author{Jesse van de Sande\altaffilmark{1}, 
Mariska Kriek\altaffilmark{2}, 
Marijn Franx\altaffilmark{1}, 
Rachel Bezanson\altaffilmark{3}, 
and Pieter G. van Dokkum\altaffilmark{4} 
}

\altaffiltext{1}{Leiden Observatory, Leiden University, P.O.\ Box
9513, 2300 RA Leiden, The Netherlands.}

\altaffiltext{2}{Astronomy Department, University of California at
Berkeley, Berkeley, CA 94720, USA}

\altaffiltext{3}{Steward Observatory, University of Arizona, Tucson,
AZ 85721, USA}

\altaffiltext{4}{Department of Astronomy, Yale University, P.O. \ Box
208101, New Haven, CT 06520-8101, USA.}


\begin{abstract}

The Fundamental Plane (FP) of early-type galaxies, relating the
effective radius, velocity dispersion, and surface brightness, has long
been recognized as a unique tool for analyzing galaxy structure and
evolution. With the discovery of distant quiescent galaxies and the
introduction of high sensitivity near-infrared spectrographs, it is
now possible to explore the FP out to $z\sim2$. In this Letter we
study the evolution of the FP out to $z\sim2$ using kinematic
measurements of massive quiescent galaxies ($M_{*}>10^{11}
M_{\odot}$). We find preliminary evidence for the existence of an FP
out to $z\sim2$.  The scatter of the FP, however, increases from
$z\sim0$ to $z\sim2$, even when taking into account the larger
measurement uncertainties at higher redshifts. We find a strong
evolution of the zero point from $z\sim2$ to
$z\sim0:\Delta\log_{10}M/L_g\propto(-0.49\pm0.03)~z$.  In order to
assess whether our spectroscopic sample is representative of the
early-type galaxy population at all redshifts, we compare their
rest-frame $g-z$ colors with those from a larger mass complete sample
of quiescent galaxies. At $z>1$ we find that the spectroscopic sample
is bluer. We use the color offsets to estimate a mass-to-light ratio
($M/L$) correction. The implied FP zero point evolution after
correction is significantly smaller:
$\Delta\log_{10}M/L_g\propto(-0.39\pm 0.02)~z$.  This is consistent
with an apparent formation redshift of
$z_{\rm{form}}=6.62^{+3.19}_{-1.44}$ for the underlying population,
ignoring the effects of progenitor bias. A more complete spectroscopic
sample is required at $z\sim2$ to properly measure the $M/L$ evolution
from the FP evolution.
\end{abstract}


\section{Introduction}
\label{sec:introduction}

The Fundamental Plane (FP) of early-type galaxies is the empirical
relation between the effective radius $r_{\rm e}$, stellar velocity
dispersion within one effective radius $\sigma_{\rm e}$, and the
average surface brightness within one effective radius $I_{\rm e}$
(e.g., \citealt{djorgovski1987}; \citealt{dressler1987};
\citealt{jorgensen1996}).  Traditionally, the offset in the FP is
interpreted as the mass-to-light ratio ($M/L$) evolution of galaxies
(e.g., \citealt{faber1987}), under the assumption that early-type
galaxies form a homologous family, and that all the evolution is
caused by the change in luminosity. \citet{holden2010} indeed suggests
that this assumption is correct, finding that out to $z\sim1$ the
slope of the FP does not change.  By extending $M/L$ evolutionary
studies to higher redshifts, it is possible to put constraints on the
formation epoch of massive galaxies.

The evolution of the $M/L$ has been studied extensively out to
$z\sim1.3$ (e.g., \citealt{vandokkum2007}; \citealt{holden2010}). The
general consensus is that the evolution of the $M/L$ appears to evolve
as $\Delta\ln~M/L_{\rm{B}}\propto~z$.  With recent studies showing
that the first massive, quiescent galaxies were already in place when
the universe was only $\sim$3 Gyr old (e.g., \citealt{kriek2006};
\citealt{williams2009}), the question arises of whether the FP already
existed at this early epoch and how the $M/L$ evolved.

With the advent of new NIR spectrographs, such as VLT-X-SHOOTER and
Keck-MOSFIRE, it is now possible to obtain rest-frame optical spectra
of quiescent galaxies out to $z\sim2$. For example, in
\citeauthor{vandesande2011} (\citeyear{vandesande2011};
\citeyear{vandesande2013}) we obtained stellar kinematic measurements
for five massive quiescent galaxies at $1.4<z<2.1$ (see also
\citealt{toft2012}; \citealt{belli2014b}).  Combined with
high-resolution imaging and multi-wavelength catalogs, these recently
acquired kinematic measurements enable the extension of FP studies
beyond $z\sim1.3$.

In this Letter, we explore the existence of an FP at $z\sim2$, and use
the FP to measure the evolution of the $M/L$ from $z\sim2$ to the
present day for massive quiescent galaxies. In a parallel study,
\citet{bezanson2013b} presented the mass FP evolution. Throughout the
Letter we assume a $\Lambda$CDM cosmology with $\Omega_\mathrm{m}$=0.3
$\Omega_{\Lambda}=0.7$, and $H_{0}=70$ km s$^{-1}$ Mpc$^{-1}$. All
broadband data are given in the AB-based photometric system.

\section{Data}
\label{sec:data}

For the work presented in this Letter, we use a variety of data sets,
which all contain accurate kinematic measurements of individual
galaxies and high quality broadband photometric catalogs. For more
details see Table \ref{tab:tab1} and J. van de Sande et al. (submitted).
All velocity dispersions presented here are aperture corrected to one 
effective radius following the method as
described in \citet{vandesande2013}.

%
\begin{deluxetable*}{l r l l l l l }[!ht]

\tabletypesize{\scriptsize}
\tablewidth{0pt}
\tablecaption{Data References Sample}

\tablehead{\colhead{Survey \& Field} & \colhead{N$_{\rm gal}$} &
\colhead{z} & \colhead{Spectroscopy} & \colhead{Telescope \& } &
\colhead{Photometric Catalog} & \colhead{Structural Parameters} \\ & &
& & \colhead{Instrument} & & \\}

\vspace{0.2cm} SDSS DR7 & 4621 & $0.05<z<0.07$ &\citet{abazajian2009}
& SDSS & \citet{blanton2005} & \citet{simard2011} \\

NMBS-COSMOS & 3 & $0.7<z<0.9$ & \citet{bezanson2013b} & Keck-DEIMOS
&\citet{skelton2014} & \citet{bezanson2011} \\ & 10 & & & &
\citet{whitaker2011} & \\

UKIDSS-UDS & 3 & $0.6<z<0.7$ & \citet{bezanson2013b} & Keck-DEIMOS&
\citet{skelton2014} & \citet{vanderwel2012} \\ \vspace{0.2cm} & 1 & &
& &\citet{williams2009} & \\

MS 1054-0321 & 8 & $z=0.83$ & \citet{wuyts2004}& Keck-LRIS &
\citeauthor{forsterschreiber2006} & \citet{blakeslee2006} \\ & & & & &
\citeyear{forsterschreiber2006} & \\ \vspace{0.2cm} GOODS-S & 7 & $
0.9 < z < 1.2 $ & \citet{vanderwel2005}& VLT-FORS2 &
\citet{skelton2014} & \citet{vanderwel2012} \\

\vspace{0.2cm} GOODS-N & 1 & $z=1.315$ & \citet{newman2010}&
Keck-LRIS & \citet{skelton2014} & \citet{vanderwel2012} \\

EGS & 8 & $1.0 < z< 1.3$ & \citet{belli2014a}& Keck-LRIS &
\citet{skelton2014} & \citet{vanderwel2012} \\ COSMOS & 6 & $1.1<z<1.3 $ 
& \citet{belli2014a}& Keck-LRIS & \citet{skelton2014} &
\citet{vanderwel2012} \\ \vspace{0.2cm} GOODS-S & 1 & $z=1.419 $ &
\citet{belli2014a} & Keck-LRIS & \citet{skelton2014} &
\citet{vanderwel2012} \\

NMBS-COSMOS & 4 & $1.2<z<1.5$ & \citet{bezanson2013a}& Keck-LRIS &
\citet{whitaker2011} & \citet{bezanson2013a}\\ \vspace{0.2cm}
NMBS-AEGIS & 2 & $1.4<z<1.6$ & \citet{bezanson2013a} & Keck-LRIS&
\citet{whitaker2011} & \citet{bezanson2013a} \\

NMBS--COSMOS & 2 & $1.6<z<2.1$ & \citet{vandesande2013} & VLT-XShooter
& \citet{skelton2014} & \citet{vandesande2013} \\ & 1 & & & &
\citet{whitaker2011} & \\ UKIDSS-UDS & 1 & $1.4<z<2.1$ &
\citet{vandesande2013} & VLT-XShooter & \citet{skelton2014} &
\citet{vandesande2013} \\ \vspace{0.2cm} &1 & & &
&\citet{williams2009} & \\

COSMOS & 1 & $z=1.823 $ & \citet{onodera2012} & Subaru-MOIRCS &
\citet{muzzin2013a} & \citet{onodera2012} \\ \vspace{0.2cm} MUSYC 1255
& 1 & $z=2.286$ & \citet{vandokkum2009}&Gemini-GNIRS &
\citet{blanc2008} & \citet{vandokkum2009} \\

COSMOS & 2 & $ 2.1 < z < 2.3 $ 
& \citet{belli2014b}& Keck-MOSFIRE & \citet{skelton2014} &
\citet{belli2014b} \\ 

\enddata
\label{tab:tab1}
\end{deluxetable*}

%
%

Stellar masses are derived using the FAST code \citep{kriek2009}. We
use the \citet{bruzual2003} Stellar Population Synthesis (SPS) models
and assume an exponentially declining star formation history, the
\citet{calzetti2000} dust attenuation law, and the
\citet{chabrier2003} stellar initial mass function.  For galaxies in
the Sloan Digital Sky Survey (SDSS), stellar masses are from the MPA-JHU
DR7\footnote{http://www.mpa-garching.mpg.de/SDSS/DR7/} release which
are based on \citet{brinchmann2004}.

The photometry and thus also the stellar mass are corrected for
missing flux using the best-fit S\'ersic luminosity
\citep{taylor2010}. Effective radii and other structural parameters,
such as \ser index and axis ratio, are determined using two-dimensional S\'ersic
fits with GALFIT \citep{peng2010}. The effective radii are
circularized, i.e., $r_{\rm{e}}=\sqrt{ab}$. All rest-frame fluxes,
including those for the SDSS sample, are calculated using the
photometric redshift code EAZY (v46; \citealt{brammer2008}).

We derive the average surface brightness within one $r_{\rm e}$
($I_{\rm{e}}$, in units of $L_{\odot,\rm{g}}~\rm{pc}^{-2}$) by
dividing the total luminosity in the rest-frame $g$-band by $2\pi
r_{\rm{e}}^2$. Absolute $g$-band magnitudes are calculated with
$M_{\odot,~\rm g}=5.14$, which we derive from the solar spectrum taken
from the CALSPEC
database\footnote{http://www.stsci.edu/hst/observatory/cdbs/calspec.html}.

We adopt a mass limit ($M_*>10^{11}~M_{\odot}$) to homogenize the
final sample.  We note, however, that our sample remains relatively
heterogeneous and in particular the higher redshift samples are biased
toward the brightest galaxies (see also Section \ref{sec:bias}).  We
furthermore use the $U-V$ versus $V-J$ rest-frame color selection
criterion to remove star-forming galaxies from our sample (e.g.,
\citealt{wuyts2007}; \citealt{williams2009}).  Quiescent galaxies are
selected to have $U-V>(V-J )\times0.88+0.59$.  This criteria is
slightly different from previous work, as we do not require that
$U-V>1.3$ or $V-J<1.5$.  The latter criteria remove post-starburst
galaxies and very old galaxies, respectively.
%
%

\begin{figure*}[!th]
\epsscale{1.18}
\plotone{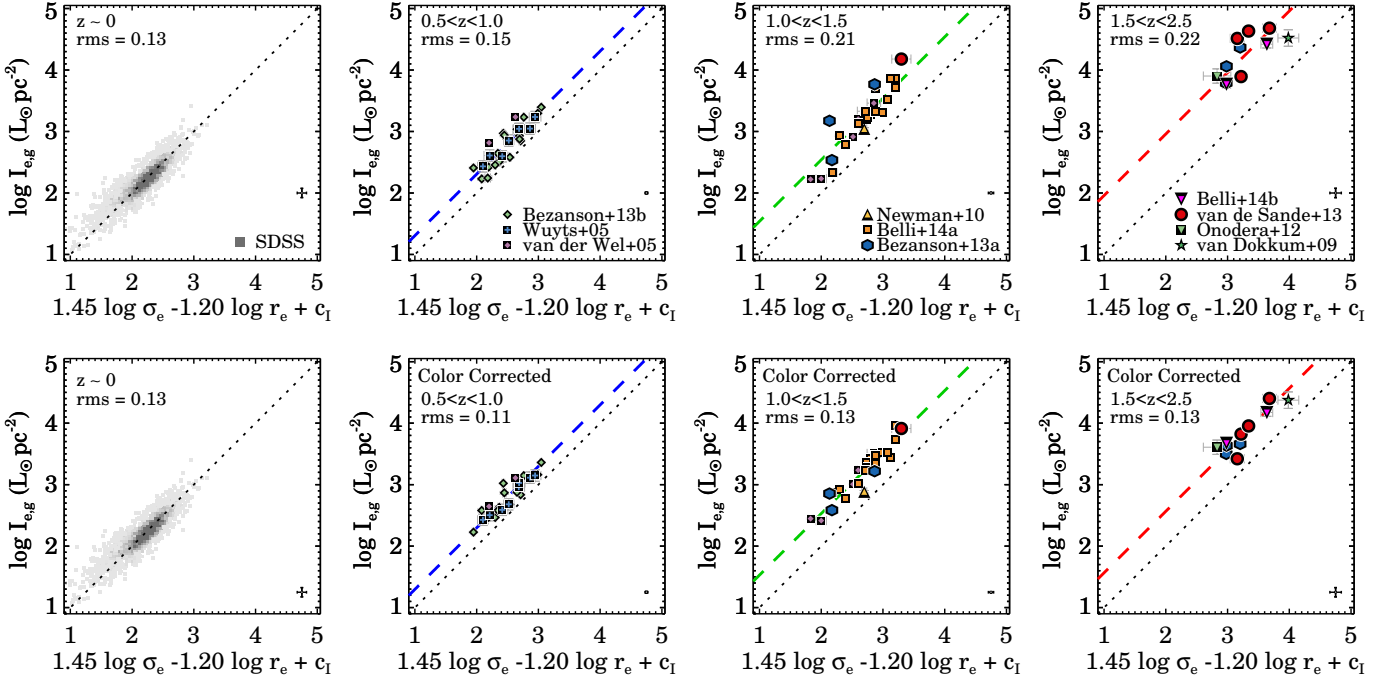}
\caption{Edge-on projections of the FP of massive quiescent galaxies
($M_{*}>10^{11}M_{\odot}$) from low to high redshift.
{Top:} at $z>0.5$ galaxies no longer fall on the SDSS FP
(dotted line), with the dashed line indicating the best-fit
relation. The rms scatter increases from $z\sim0-z\sim2$, and
cannot be explained by larger measurement errors at $z>1$ alone.
{Bottom:} similar as top row, but now including a color
correction applied to all galaxies in our spectroscopic sample, to
fix the sample bias toward bright young galaxies (see Section
\ref{sec:bias}). With the correction applied, we find that the scatter
around the FP now reduces for the three high-redshift bins, implying
that a tight FP exists out to $z\sim2$.}
\label{fig:fig1}
\end{figure*}
%

\section{The Fundamental Plane}
\label{sec:fp}

The FP is traditionally  written as: 
\begin{equation}
\log_{10}r_{\rm{e}}=a\log_{10}\sigma_{\rm{e}}+b\log_{10}I_{\rm{e,g}}+c_{\rm{r}}
\label{eq:fpre}
\end{equation}
with $r_{\rm{e}}$ in kpc, $\sigma_{\rm{e}}$ in km\,s$^{-1}$, and
$I_{\rm{e,g}}$ in $L_{\odot,\rm{g}}~\rm{pc}^{-2}$. In this Letter, we
adopt the slope from \citet{jorgensen1996}, i.e., $a=1.20$, and
$b=-0.83$.  While detailed studies on galaxies in the SDSS have shown
that the slope is steeper, i.e., $a=1.404$, and $b=-0.76$
\citep{hyde2009}, we nonetheless adopt the \citet{jorgensen1996}
values, for an easier comparison with previous high-redshift
studies. We do not fit the slope ourselves at high redshift, as our
sample is too small and biased at $z>1$. However, we note that
\citet{holden2010} find the same slope of the FP at $z\sim1$ as
\citet{jorgensen1996}.

While the projection of the FP along the effective radius is most
often shown, the projection along $\log_{10}I_{\rm{e}}$ directly
shows the evolution in the $M/L$, which is thought to be the main
driver in the evolution of the FP zero point. The top row in Figure
\ref{fig:fig1} shows the following projection:
\begin{equation}
\log_{10}I_{\rm{e}}=a\log_{10}\sigma_{\rm{e}}+b\log_{10}r_{\rm{e}}+c_{\rm{I}},
\label{eq:fpie}
\end{equation}
with $a=1.45$, $b=-1.20$, and $c_{\rm I}=-0.11$. We determine the zero
point $c_{\rm I}$ using a least-square fit, using the $IDL$ function
$MPFIT$ \citep{markwardt2009}. The zero point evolves rapidly with
redshift: $0.30,~0.54,~0.96~\rm{dex},~\rm{at}~
z\sim0.75,~z\sim1.25,~z\sim2.00$, respectively.

The top row in Figure \ref{fig:fig1} suggests that the FP exists out to
$z\sim2$. However, to quantify the existence of an FP, we consider the
scatter before and after the FP fit, i.e., the rms scatter in
$\log_{10}I_{\rm{e}}$ versus the scatter in $(a
\log_{10}\sigma_{\rm{e}}+b\log_{10}r_{\rm{e}})-\log_{10}I_{\rm{e}} $
or $\Delta\log_{10}I_{\rm{e}}$.
For example, for galaxies in the SDSS for which we know the FP exists,
the scatter in $\log_{10}I_{\rm{e}}$ is $0.26\pm0.003$ dex before the
fit, while after the fit, we measure
$\Delta\log_{10}I_{\rm{e}}=0.13\pm0.002$dex.  At $z=2$ we find a
decrease as well, from $\log_{10}I_{\rm{e}}=0.33\pm0.04$ dex, before
the fit, to $\Delta\log_{10}I_{\rm{e}}=0.25\pm0.04$ after the
fit. This hints at the existence of an FP at $z\sim2$, although the
scatter in dex is almost twice as high compared to $z\sim0$.

We use Monte Carlo simulations to determine if the scatter could be
due to measurement uncertainties alone. We find that scatter induced
by observational errors is $0.14~$dex at $z\sim2$, resulting in an
intrinsic scatter of $0.20~$dex. Compared to the FP scatter at
$z\sim0$, the intrinsic scatter at $z\sim2$ is still higher. The
larger scatter could have been caused by the large range in age and
the bias toward post-starburst galaxies.

\begin{figure*}
\epsscale{1.18}
\plotone{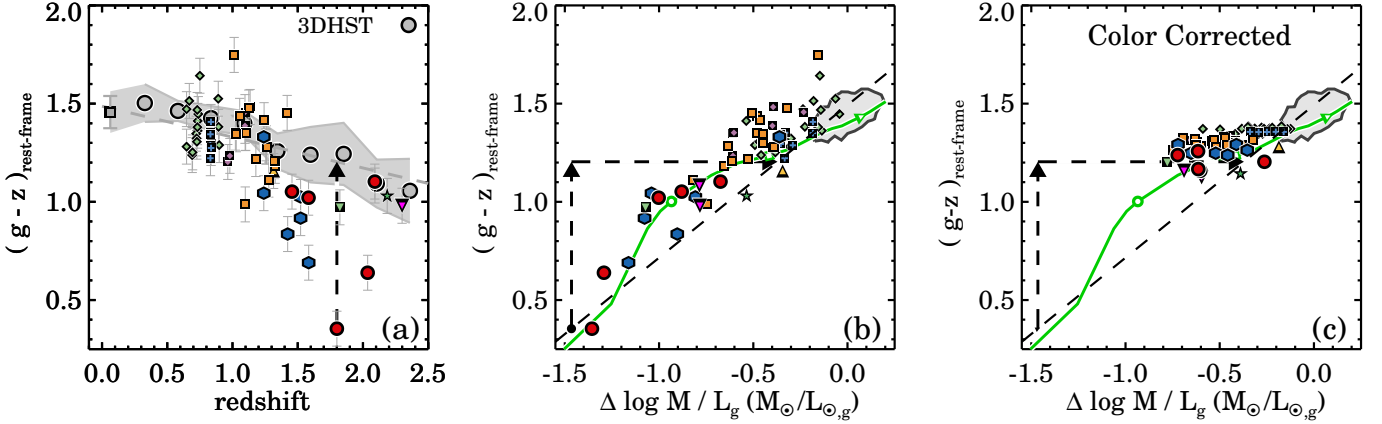}
\caption{
Redshift evolution of rest-frame $g-z$ color and $M/L$ correction from
color offset.  (a) $(g-z)_{\rm{rest-frame}}$ colors for our
high-redshift spectroscopic sample compared to a larger and more
complete photometric sample from the 3DHST survey with a similar mass
selection as shown by the large gray circles. The gray area indicates
the $1-\sigma$ scatter. For the reference sample we measure that
$(g-z)_{\rm{\rm{rest-frame}}}$ evolves as $-0.16z$ (dashed gray
line). At $z>1$ the spectroscopic sample becomes increasingly bluer
than the reference sample, indicating a rather strong selection bias.
(b) $\log_{10}{M/L_g}$ vs. rest-frame $g-z$ color. Massive galaxies
from the SDSS are shown by the gray contour, which encloses 68\% of
all galaxies. The green line shows an Ma11 model, and the dashed black
line shows the linear approximation to the model
($\log_{10}M/L\propto1.29(g-z)_{\rm{rest-frame}}$). We use the model
approximation to correct the $M/L$ using the color offset. The
correction is demonstrated with arrows for the galaxy with the
lowest $M/L$ and bluest color. (c) Corrected $M/L$s and $g-z$
colors. The correction is small for most low-redshift galaxies
($\sim0.2$dex), while the galaxies at $z>1.5$ have larger corrections
of $\sim0.2-1.0$dex.
}
\label{fig:fig2}
\end{figure*}
%
%
%
\section{Correcting for Sample Bias}
\label{sec:bias}

In order to explain the larger scatter in the FP of our high-redshift
sample, we compare the rest-frame $g-z$ colors of our spectroscopic
sample to a larger photometric and more complete galaxy sample. We use
galaxies from the 3D-HST survey (v4.1; \citealt{brammer2012};
\citealt{skelton2014}), which were selected to be massive
($M_{*}>10^{11}M_{\odot}$), non-star-forming according to their $U-V$
versus $V-J$ colors (see Section \ref{sec:data}), and have a reduced
$\chi^2<5$ for the best-fit stellar population model to their
spectral energy distribution. Rest-frame fluxes for the reference sample were derived with
EAZY, in a similar fashion as for the spectroscopic sample and are
based on photometric redshifts.

In Figure \ref{fig:fig2}(a) we show the comparison of the rest-frame
$g-z$ color as a function of redshift. The reference sample is binned
in redshift, with the median of each bin shown as a big gray
circle. The dashed line shows the best fit of the color as a function
of redshift: $(g-z)_{\rm{rf}}=-0.16z+1.49 $. There is very little
difference between the reference and the spectroscopic sample at
$0<z<1$, while at $z>1$ the spectroscopic sample becomes increasingly
bluer compared to the median of the reference sample. At $z>1.4$ we
furthermore see a large spread in colors, where some galaxies follow
the median of the galaxy population, while other galaxies are offset
by more than 0.5 mag.

All spectroscopic samples are relatively blue at higher redshifts, as
it is easier to obtain high signal-to-noise spectra for quiescent
galaxies compared to the much fainter older quiescent galaxies.  In
order to estimate the effect of this bias, we use the color difference
with the reference sample to calculate a $M/L$ correction for the
individual galaxies in our spectroscopic sample.

First, we measure the relative evolution of the $M/L$ from the FP as
shown in the top panels of Figure \ref{fig:fig1}, under the assumption
that $a$, and $b$ do not evolve with redshift:
\begin{equation}
\Delta\log_{10}M/L=c_{\rm{z}}-c_{\rm{0}}.
\label{eq:deltaml}
\end{equation}
Here $c_{\rm{0}}$ is the FP zeropoint at redshift zero, while
$c_{\rm{z}}$ is determined from the residuals from the FP for each
individual galaxy at redshift $z:$
\begin{equation}
c_{\rm{z}}=(a\log_{10}\sigma_{\rm{e}}+b\log_{10}r_{\rm{e}})-\log_{10}I_{\rm{e}}
\label{eq:czi}
\end{equation}
Note that we have $\Delta\log_{10}M/L=\Delta\log_{10}I_e$.

Next, we use the relation between the rest-frame $g-z$ color and the
derived $\Delta\log_{10}M/L$ as shown in Figure \ref{fig:fig2}(b) to
derive the $M/L$ correction.  There is a correlation between the color
and $M/L$, where the bluest galaxies have the lowest $M/L$.  This
correlation is also predicted by SPS models, on which we base our
$M/L$ correction. In Figure \ref{fig:fig2}(b) we show a
\citeauthor{maraston2011} (\citeyear{maraston2011}, Ma11) solar
metallicity model (green line), with a truncated star formation
history and constant star formation for the first 0.5-Gyr. Different
ages in this model are indicated on the green line by the circle
(0.3-Gyr), diamond (1.0-Gyr), and triangle (10-Gyr). We approximate
the Ma11 model by a simple linear fit:
$\log_{10}M/L\propto1.29(g-z)_{\rm{rf}}$ (dashed black line). We use
this fit to estimate the $M/L$ correction.

As an example, we highlight the bluest galaxy in our sample,
NMBS-Cos-7447 at $z\sim1.8$.  In Figure \ref{fig:fig2}(a), the black
arrow indicates the color offset to the $(g-z)_{\rm{rest-frame}}$
color-redshift relation. The same galaxy has an extremely low $M/L$ as
is clear from Figure \ref{fig:fig2}(b). Here the vertical arrow is again
the color offset of this galaxy, but the horizontal arrow now points
toward the $M/L$ that the galaxy would have, if it would fall on the
color-redshift relation from Figure \ref{fig:fig2}(a). The $M/L$
correction that we apply to this individual galaxy is therefore the
length of the horizontal arrow. In other words, for each galaxy we
lower the $M/L$ by 1.29 times the color offset, where the factor 1.29
comes from the linear fit to the Ma11 model (dashed line).  We note,
however, that a steeper $M/L$ versus color relation yields larger $M/L$
corrections, and thus impacts the $M/L$ evolution in Section
\ref{sec:mlevo}.  We show the corrected colors and $M/L$ for the
entire sample in Figure \ref{fig:fig2}(c). The rest-frame $g-z$ color
and $M/L$ of the spectroscopic sample is now similar to that of the
reference sample.

%


%
\begin{figure*}
\epsscale{1.0}
\plotone{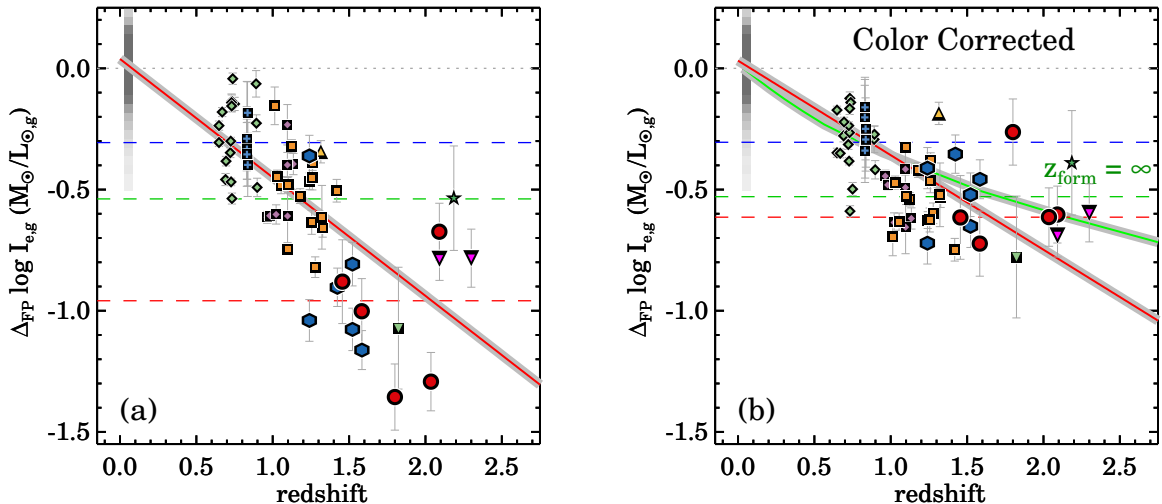}
\caption{(a) Redshift evolution of the residuals from the FP. We show
the residual from the FP, $\Delta\log_{10}I_{\rm{e,g}}$ vs.
redshift as compared to $z\sim0$ , with the dashed colored lines
corresponding to the best-fit offset for the different redshift
regimes indicated in Figure \ref{fig:fig1}. The red line shows the
best fit described by $\Delta\log_{10}M/L\propto-0.49z\pm0.03$,
where we have anchored the fit to the median of the SDSS data. We find
a similar evolution of the $M/L$ compared to previous work, while
the high-redshift data at $z>1.25$ have lower $M/L$. (b) Similar to
(a), but now including a correction based on the offset from the
$(g-z)_{\rm{rest-frame}}$ relation. This time, we find a milder
evolution of $\Delta\log_{10}M/L_{\rm{g}}\propto(-0.39\pm0.02)~z$
indicated by the red line, while we show the uncorrected evolution as
the dashed gray line. We also include a single burst Ma11 model, with
solar metallicity, and with $z_{\rm{form}}=\infty$ (green line).
}
\label{fig:fig3}
\end{figure*}


%
With the $M/L$ (or $\Delta\log_{10}I_{\rm{e}}$) correction applied, we
can return to the FP which is shown in the bottom row of Figure
\ref{fig:fig1}. As before, the dashed colored lines indicate the
best-fit FP with the following zero points in each redshift bin: 0.30,
0.53, and 0.62 dex, from low to high redshift.  Note that the $z\sim2$
zero point is significantly lower than before the correction. The
scatter around the FP is greatly reduced for all three redshift
bins. In particular for the highest redshift bin, which had the
largest spread in $(g-z)_{\rm rf}$ color, the scatter has decreased by
almost a factor two. More quantitatively, the scatter after the FP fit
is: $\Delta\log_{10}I_{\rm{e}}=0.11\pm0.02,~0.13\pm0.02,~0.14\pm0.02$,
from low to high redshift, respectively. We note that the small
scatter is most likely a lower limit, as we do not include the
intrinsic color scatter in our analysis.

\section{Evolution of the $M/L$}
\label{sec:mlevo}
In Figure \ref{fig:fig3} we show the $M/L$ as derived from the FP
offset relative to $z\sim0$, as a function of redshift for all
galaxies. We indicate the different zero points from Figure
\ref{fig:fig1} as dashed lines. The red solid line is a linear
least-square fit to all data, which describes the evolution of the
$M/L:\Delta\log_{10}M/L_{\rm{g}}\propto(-0.49\pm0.03)~z$.  Here we
have anchored the fit to the median of the $z\sim0$ SDSS
galaxies. When excluding SDSS from the fit, the evolution is more
rapid:~$\Delta\log_{10}M/L_{\rm{g}}\propto(-0.61\pm 0.09)~z$.

The evolution is similar to a previous result by
\citet{vandokkum2007}, who find that the
$M/L\propto(-0.555\pm0.042)~z$ in the rest-frame $B$ band. Here, the
effect of using a slightly bluer rest-frame filter makes very little
difference, i.e., our measured $M/L$ evolution in the rest-frame
$B$ band is only slightly faster as compared to the $g$ band:
$\Delta\log_{10}M/L_{\rm B}\propto(-0.53\pm0.03)~z$.

For a fair comparison with previous studies, we also restrict the fit
to $z<1.4$, and find a slower evolution than other studies:
$~\Delta\log_{10}M/L_{\rm{g}}\propto(-0.45\pm0.01)~z$.  This suggests
that the relatively fast evolution of the entire sample is mainly
driven by galaxies at $z>1.4$. This is furthermore evident from Figure
\ref{fig:fig3}(a) in which almost all $z>1.4$ galaxies fall below the
best fit for the entire sample.

Figure \ref{fig:fig3}(b) shows the $M/L$ evolution with redshift, with
the $M/L$ correction included. This time, we find a milder evolution
in the $M/L$ than before:
$\Delta\log_{10}M/L_{\rm{g}}\propto(-0.39\pm0.02)~z$, shown by the red
line.  Our corrected $M/L$ evolution is similar to previous studies by
e.g., \citet{vandokkum2003} ($-0.460\pm0.039$), and \citet{holden2005}
($-0.426\pm0.04$), but slower than the evolution found by
\citet{vandokkum2007} ($-0.555\pm0.04$), and \citet{holden2010}
($-0.60\pm0.04$).  If we would have used a steeper $M/L$ vs. color
relation for the correction,
(e.g.,$~\log_{10}M/L\propto~1.80(g-z)_{\rm{rest-frame}}$) the $M/L$
evolution would be
slower$:~\Delta\log_{10}M/L_{\rm{g}}\propto(-0.36\pm0.03)~z$.

We also show the single burst Ma11 model from Figure \ref{fig:fig2},
with a formation redshift of infinity. Using the same Ma11 model, we
fit the average formation redshift of the biased sample (Figure
\ref{fig:fig3}(a)), and estimate
$z_{\rm{form}}=3.29^{+0.55}_{-0.56}$. For our corrected sample (Figure
\ref{fig:fig3}(b)) we estimate the apparent
$z_{\rm{form}}=6.62^{+3.19}_{-1.44}$. The large difference highlights
the strong selection bias in the sample. Both estimates ignore the
progenitor bias, which is very strong given the fact that the number
density of massive galaxies changes by a factor of
$\sim10~\rm{from}~z\sim2~\rm{to}~z\sim0$
\citep{muzzin2013b}.Therefore, the corrected $z_{\rm{form}}$ should be
regarded as an upper limit for a mass-complete sample that is not
supplemented by newly quenched galaxies at later times. We note that
we would find a lower $z_{\rm{form}}$ if we were to include progenitor
bias in our analysis, or if we would use an evolving mass limit in our
sample-selection. A more detailed study of this effect would require a
mass-complete spectroscopic sample.

\section{Discussion}
\label{sec:discussion}

In Section \ref{sec:bias} we have corrected our sample for the bias
toward young quiescent galaxies. In this correction we assume that all
quiescent galaxies at a particular redshift have the same color and
thus the same age. This assumption is an oversimplification, and the
scatter in the FP is actually partly due to age variations (e.g.,
\citealt{graves2009}). Furthermore, the age variations of quiescent
galaxies may increase with redshift, as shown by
\citet{whitaker2010}. Thus, it is very likely that the scatter in the
FP will be induced by larger age variations at higher redshifts.

Because we apply the same correction at all redshifts, we do show
that the scatter in the FP due to variation in properties other than
age (e.g., evolution in stellar mass, size, velocity dispersion, and
metallicity) is approximately constant over time. However, our $M/L$
evolution is not necessarily the $M/L$ evolution of progenitors and
descendants of a fixed population, as the masses, sizes, and velocity
dispersions of these galaxies likely evolve systematically with
redshift.
Using a simple model, we can estimate the effect of structural
evolution on the FP and $M/L$ evolution.  Given that mass evolves as
$\Delta\log_{10}M/M_{\odot}\sim0.15z$ \citep{vandokkum2010}, we found
that
$\Delta~r_e\propto~M^{1.83}\simeq~M^2~,~\rm{and}~\Delta~\sigma_e\propto~M^{-0.49}\simeq~M^{-0.5}$
in \citet{vandesande2013}.  Thus, if mass grows by a factor $f$, the
$M/L$ evolution (from Equations (\ref{eq:deltaml}) and (\ref{eq:czi}))
including structural evolution can be written as follows:
\begin{equation}
\Delta\log_{10}M/L\simeq~a\log_{10}(f^{-0.5}\sigma_{\rm{e}})+b\log_{10}(f^{2}r_{\rm{e}})-\log_{10}(f^{-3}I_{\rm{e}})
\label{eq:mlscatter}
\end{equation}
The bias in the $M/L$ evolution induced by structural evolution can
therefore be approximated by $(-0.5a+2b+3)\log_{10}f$. Assuming a mass
growth factor of 0.3~dex, we find that the offset in the $M/L$ due to
structural evolution is small: $\sim0.04~\rm{dex}$.  We note that if
velocity dispersion does not evolve, this effect is larger: $\sim0.2$
dex.

\section{Conclusion}
\label{sec:conclusions}

In this Letter, we have used stellar kinematic and structural
measurements of massive quiescent galaxies ($M_*>10^{11}M_{\odot}$)
out to $z\sim2$ to study the evolution of the rest-frame $g$-band
FP. We utilize this empirical relation between the
size, stellar velocity dispersion, and the surface brightness, to
constrain the evolution of the $M/L$.

We find preliminary evidence for the existence of an FP out to
$z\sim2$, but with larger scatter as compared to the present-day FP
for massive quiescent galaxies from the SDSS. There is a rapid
evolution of the FP zero point from $z\sim0$ to
$z\sim2:\Delta\log_{10}M/L_{\rm{g}}\propto(-0.49\pm0.03)~z$.
Furthermore, we find that the $M/L_{\rm{g}}$ evolution for galaxies at
$z>1.4$ is faster than for galaxies at $z<1.4$.

The larger scatter and fast evolution can be explained by the fact
that our spectroscopic sample becomes increasingly bluer at high
redshift compared to a mass complete sample of quiescent galaxies. We
calculate the color difference between the galaxies and the mass
complete sample, and estimate the systematic effect on the $M/L$.

With this correction applied, the evolution of the $M/L$, as derived
from the FP, is slower:
$\Delta\log_{10}M/L_{\rm{g}}\propto(-0.39\pm0.02)~z$.  A simple model,
ignoring progenitor bias, would imply a formation redshift of
$z_{\rm{form}}=6.62^{+3.19}_{-1.44}$ for a mass complete sample. The
difference between the evolution of our observed sample and the
underlying population, highlights the need for a more detailed study
of a mass complete spectroscopic sample.

\acknowledgments{
We thank the anonymous referee for constructive comments. We thank the
NMBS and 3DHST collaborations, and thank Rik Williams, Ryan Quadri,
and Andrew Newman for providing ancillary data. This work is based on
observations taken by the 3D-HST Treasury Program (GO-12177,12328)
with the NASA/ESA-HST, which is operated by the Association of
Universities for Research in Astronomy, Inc., under NASA contract
NAS5-26555.}


\clearpage

\end{document}